\documentclass[12pt,preprint]{aastex}
\usepackage{pdflscape}
\usepackage{color, soul}
\usepackage{booktabs}
\usepackage{threeparttable}
\usepackage{multirow}
\usepackage{ulem}

\slugcomment{}

\begin{document}
\setstcolor{red}


\title{Very Large Array multi-band monitoring observations of M\,31*}

\author{Yang Yang\altaffilmark{1,2}, Zhiyuan Li\altaffilmark{1,2}, Lor\'{a}nt O. Sjouwerman\altaffilmark{3}, Feng Yuan\altaffilmark{4,5}, Zhi-Qiang Shen\altaffilmark{4,6}}
\altaffiltext{1}{School of Astronomy and Space Science, Nanjing University, Nanjing 210023, China}
\altaffiltext{2}{Key Laboratory of Modern Astronomy and Astrophysics, Nanjing University, Nanjing 210023, China}
\altaffiltext{3}{National Radio Astronomy Observatory, Socorro, NM 87801, USA}
\altaffiltext{4}{Shanghai Astronomical Observatory, Chinese Academy of Sciences, Shanghai 200030, China}
\altaffiltext{5}{Key Laboratory for Research in Galaxies and Cosmology, Chinese Academy of Sciences, China}
\altaffiltext{6}{Key Laboratory of Radio Astronomy, Chinese Academy of Sciences, China}
\email{lizy@nju.edu.cn}



\begin{abstract}
The Andromeda galaxy (M\,31) hosts one of the nearest and most quiescent super-massive black holes, which provides a rare, but promising opportunity for studying the physics of black hole accretion at the lowest state.
We have conducted a multi-frequency, multi-epoch observing campaign, using the Karl G. Jansky Very Large Array (VLA) in its extended configurations in 2011-2012, to advance our knowledge of the still poorly known radio properties of M\,31*.
For the first time, we detect M\,31* at 10, 15 and 20 GHz, and measure its spectral index, $\alpha \approx -0.45\pm0.08$ (S$_{\nu}$ $\varpropto$ $\nu^{\alpha}$), over the frequency range of 5-20 GHz. 
The relatively steep spectrum suggests that the observed radio flux is dominated by the optically-thin part of a putative jet, which is located at no more than a few thousand Schwarzschild radii from the black hole. On the other hand, our sensitive radio images show little evidence for an extended component, perhaps except for several parsec-scale ``plumes", the nature of which remains unclear.
Our data also reveal significant (a few tens of percent) flux variation of M\,31* at 6 GHz, on timescales of hours to days. 
Furthermore, a curious decrease of the mean flux density, by $\sim$50\%, is found between VLA observations taken during 2002-2005 and our new observations, which might be associated with a substantial increase in the mean X-ray flux of M\,31* starting in 2006. 
\end{abstract}


\section{Introduction}
The advent of multi-wavelength surveys in the past two decades has led to the consensus that the vast majority of super-massive black holes (SMBHs) in the local Universe are radiatively quiescent (e.g., \citealp{1997ApJ...487..568H,2005A&A...435..521N,zha09}), with bolometric luminosities only a small fraction of their Eddington limit ($\lesssim10^{-3}L_{\rm Edd}$). 
Direct probes of these so-called low-luminosity active galactic nuclei (LLAGNs; see review by \citealp{2008ARA&A..46..475H}) prove to be challenging and often require high-resolution, high-sensitivity observations. 
Nevertheless, studies of LLAGNs are crucial for our comprehensive understanding of the physics of SMBH accretion and feedback over cosmic time \citep{2012ARA&A..50..455F, 2013ARA&A..51..511K}.

Lying at the extreme faint end of the LLAGN family is the Galactic center black hole, Sgr A* \citep{2001ARA&A..39..309M}. At the radio band, in which it was first discovered, Sgr A* is a compact synchrotron source with an inverted spectrum extending into the millimeter wavelengths (e.g., \citealp{1998ApJ...499..731F}), and exhibits moderate flux variations on timescales from hours to years \citep{2003ANS...324..355Z,2004AJ....127.3399H,2011ApJ...729...44Y}.
The variability of Sgr A* is much stronger at higher frequencies \citep{2010RvMP...82.3121G}. In particular, X-ray flares have been detected on average once per day, with amplitudes reaching $\sim$100 times the quiescent level \citep{2013ApJ...774...42N}.  
Despite this remarkable (and still not well understood) variability, Sgr A* is exceptionally underluminous, showing a bolometric luminosity only $\sim 10^{-8}L_{\rm Edd}$.
Nevertheless, Sgr A* occupies a key position in the development of theories for LLAGNs.
It is now widely thought that LLAGNs are powered by a radiatively inefficient, advection-dominated accretion flow (ADAF), 
which is probably coupled with outflows in the form of jets and/or winds (see review by Yuan \& Narayan 2014). 
This ADAF-jet paradigm gains support from radio interferometric surveys of nearby galactic nuclei, in which compact cores with high brightness temperatures, and sometimes with elongated components, are frequently detected and best interpreted as synchrotron radiation from the magnetized, relativistic jets (e.g., Nagar et al.~2000, 2005).  
Ironically, the ADAF-jet paradigm is not without controversy with Sgr A* itself. 
Despite its virtue of proximity, Sgr A* is persistently seen as a compact source (i.e., detailed morphology not detected)
under the currently best available resolution afforded by VLBI observations, 
down to the vicinity of its presumed event horizon  (Bower et al.~2004; Shen et al.~2005; Doeleman et al.~2008), thus leaving serious doubt on the existence of the putative jet (but see Li et al.~2013).

The Andromeda galaxy (M\,31) hosts the second nearest SMBH, known as M\,31*, which has a dynamical mass of 1.4$^{+0.9}_{-0.3}\times10^8{\rm~M_{\sun}}$ inferred from stellar kinematics \citep{2005ApJ...631..280B}. 
M\,31* has been detected only in the radio and X-rays to date. Crane et al.~(1992) first identified a compact 8.4 GHz source at the nucleus of M\,31 using the Very Large Array (VLA).
Its physical association with the putative SMBH was reinforced by a follow-up 8.4 GHz observation that found mild variability ($\sim 30-40~\mu$Jy; \citealp{1993ApJ...417L..61C}). 
Due to its intrinsic dimness and contamination from neighboring X-ray sources, the X-ray counterpart of M\,31* was not firmly established until a much later time (Garcia et al.~2010).
The bolometric luminosity of M\,31* is estimated to be $\sim 10^{-8}L_{\rm Edd}$, making it only the second known LLAGN, after Sgr A*, to exhibit such a low state.
Interestingly, M\,31* has exhibited flaring X-ray emission since early 2006, with amplitudes similar to those seen in Sgr A* (Li et al.~2011).

The proximity of M\,31*, as well as its similarities with Sgr A*, holds promise for exploring the still poorly understood physics of SMBHs accreting at very sub-Eddington rates. 
We have carried out a systematic observing campaign of M\,31*, using the Karl G. Jansky Very Large Array (JVLA\footnote{The official acronym for the Karl G. Jansky Very Large Array is VLA. Here and in the following we use the acronym ``JVLA" to avoid potential confusion with the observations taken in the pre-JVLA era.}) in multi-epoch and multi-frequency. 
These observations of unprecedented sensitivity and resolution allow us to detect M\,31* at high frequencies, probe any extended component, constrain its instantaneous spectrum, and study its variability on various timescales. 
We describe the observations and data reduction procedure in Section 2. The results are presented in Section 3, followed by a discussion in Section 4 and a summary in Section 5. 
We adopt a distance of 780 kpc for M\,31* \citep{1998ApJ...503L.131S}. 
Thus $1^{\prime\prime}$ corresponds to 3.8 pc, or $\sim$$2.8\times10^5 R_{\rm Sch}$ (the Schwarzschild radius for a black hole mass of 1.4$\times10^8{\rm~M_{\sun}}$).
Throughout this work, errors are quoted at the 1\,$\sigma$ confidence level, unless otherwise noted.

\section{Observations and Data Reduction} \label{sec:data}
\subsection{JVLA Observations} \label{subsec:JVLA}
We have observed M\,31* with JVLA at the C, X, Ku and K bands, in a total of 19 epochs between May 2011 and December 2012 (Project ID: 11A-178, SD0487, 11A-137 and 12B-002).  
The array configurations included A, B and BnA, which are well suited for detecting a compact source. 
The central frequencies were 5.0, 6.0 or 6.6 GHz for C-band, 8.5 or 10 GHz for X-band, 15 GHz for Ku-band and 20 GHz for K-band.
We note that these observations were carried out during several stages of commissioning of the new receiver systems, hence the different projects have different parameters (especially bandwidth and central frequency).
 In 17 of the 19 epochs, a dual-frequency was employed, including C-band and one of the other three higher-frequency bands, which allows us to measure an instantaneous spectral index. The remaining two epochs were taken with the C-band only.
In total, we obtained 19 C-band, 7 X-band, 2 Ku-band and 8 K-band observations, which together provide a good sampling of the potential variability in M\,31* on timescales ranging from hours to weeks and months. 
Table \ref{tab:VLAresults} presents details of these observations (each assigned an ID), including observing date, central frequency, integration time, array configuration, bandwidth, beam size and RMS noise. 

In all observations, the phase center was placed at the nucleus of M\,31, i.e., $[RA, DEC]$ (J2000) = [00$^{h}$42$^{m}$44\fs329, +41\arcdeg16\arcmin08\farcs42].
We used 3C48 as the flux density calibrator for the amplitude scale and instrumental bandpass, with the exception of Project 11A-137, in which 3C147 was used. 
The Perley-Butler 2010 flux density scale was applied. 
The on-target scans, 6-10 minutes each, were sandwiched between scans of the nearby gain calibrator J0038+4137. 
Since the flux density of M\,31* is too weak to be self-calibrated, we relied on the cycle time being short enough to properly compensate for the temporal atmospheric phase fluctuations.

Each individual visibility data was flagged, calibrated, imaged and restored using the Common Astronomy Software Applications package\footnote{http://casa.nrao.edu} (CASA, version 4.3) in standard procedures.
We inspected all visibility data by eye, and manually flagged bad data (from RFI and instrumental effects). 
The Stokes I images were then produced using the CLEAN task, with the Multi-frequency Synthesis mode, nterms=1 and natural weighting. 
The flux densities of M\,31*, whenever detected, were measured by the IMFIT task, fitting an elliptical Gaussian to the Stokes I image. 

To maximize the signal-to-noise ratio, we used the CONCAT task to combine the individual visibility data of same central frequency into a concatenated visibility data.
The CLEAN task was employed again to make Stokes I images at different bands.
Details of the concatenated images, including central frequency, total integration time, beam size, RMS level and the individual observations involved, are given in Table \ref{tab:concat}. 

We note that Projects 11A-178 and SD0487 were taken in coordination with simultaneous {\it Chandra} observations. The relation between the radio and X-ray emission from M\,31* will be studied in a forthcoming paper.

\subsection{Historical VLA observations}
To extend our temporal baseline for variability study, 
we revisit historical VLA observations of M\,31*, conducted by one of us (L.O.S.) during 2002--2005. 
These ten observations, taken in the C-band (5 GHz) under A- or B-configuration, are summarized in Table \ref{tab:VLAresults} (IDs 37-46).
We reduced the data also using CASA, and following the standard ``old" VLA tutorial\footnote{https://casaguides.nrao.edu/index.php/Calibrating\_a\_VLA\_5\_GHz\_continuum\_survey}, to minimize systematics when compared to the JVLA data. The primary flux density calibrator was 3C48 and the gain calibrator was J0038+4137.
All Stokes I images were synthesized with natural weighting and fitted with the IMFIT task. We also used the CONCAT task to combine the ten observations to obtain a concatenated 5 GHz image.

\section{Results} \label{sec:result}
\subsection{Morphology} \label{subsec:morphology}
\begin{figure*}\centering
\epsscale{1}
\plotone{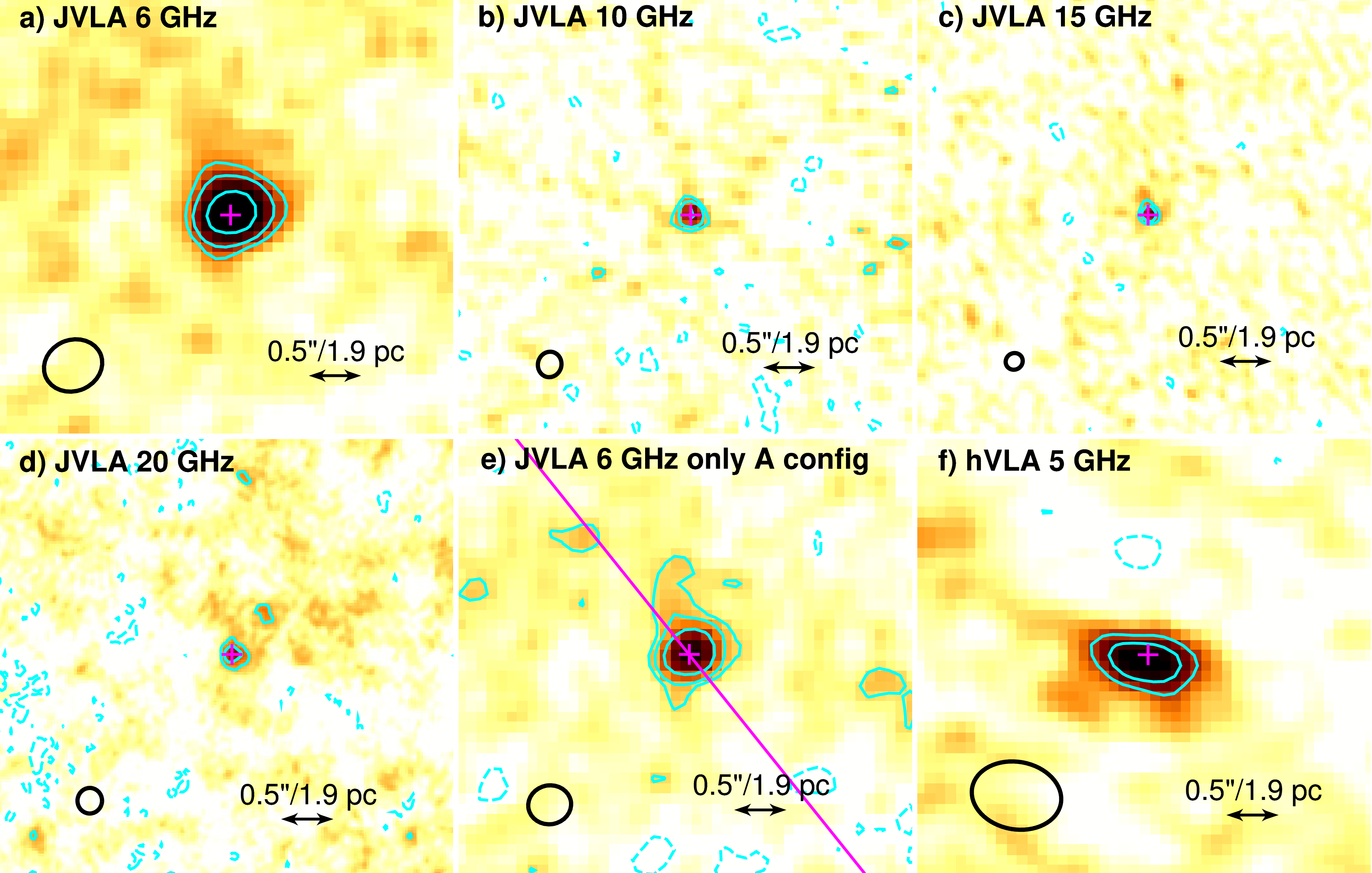}
\caption{The concatenated images of M\,31* in different bands. (a) 6 GHz; (b) 10 GHz; (c) 15 GHz; (d) 20 GHz; 
(e) 6 GHz, from only the A-array observations; (f) 5 GHz, from the historical VLA observations taken during 2002-2005.
Contour levels are at [-2, 3, 5, 10] $\times$ RMS in (a), (b) and (e), and at [-2, 3, 5] $\times$ RMS in (c), (d) and (f). 
Positive and negative levels are denoted by solid and dash contours, respectively. 
In each panel, the magenta cross marks the putative position of M\,31*, $[RA, DEC]$(J2000) = [00$^{h}$42$^{m}$44\fs325, +41\arcdeg16\arcmin08\farcs43]), 
and the ellipse at the lower left corner represents the synthesized beam.
The major axis of the M\,31 disk is indicated by a solid line (position angle of $38 \degr$) in (e).
}
\label{fig:combinedimage}
\end{figure*}

Figure \ref{fig:combinedimage}a-d display the concatenated JVLA images in the four bands, enclosing the central $4\arcsec\times4\arcsec$ ($\sim$15.2 pc$\times$15.2 pc) region of M\,31.
These images have RMS noise of 1.9, 2.0, 2.0 and 2.0 $\mu$Jy at 6, 10, 15 and 20 GHz, respectively,
and corresponding beam size of $0\farcs58\times0\farcs50$, $0\farcs25\times0\farcs23$, $0\farcs17\times0\farcs16$ and $0\farcs25\times0\farcs24$. 
M\,31* is detected as a compact source at its putative position at all four frequencies, in particular, at 10, 15 and 20 GHz for the first time. 
On the other hand, M\,31* was not detected in either of the two observations at 8.5 GHz, nor in their concatenated image (hence not shown here).

A closer look at the 6 GHz image (Figure \ref{fig:combinedimage}a) indicates substructures on a scale of $\lesssim 1^{\prime\prime}$, immediately around an otherwise point-like source. Two small ``plumes'' seem to protrude toward the north and the south, reminiscent of a bipolar outflow or jets from the central black hole. Complicating this picture, another ``plume" reaching out to the west is also hinted. 
We further show in Figure \ref{fig:combinedimage}e the concatenated C-band image including the 11 A-array JVLA observations only. 
The above substructures persist in this image of higher resolution (beam size of $0\farcs51\times0\farcs42$).
This remains true in further tests in which a subset of data is taken out from the concatenated image. 
A similar morphology appears in the 15 GHz image, but the source is virtually unresolved in the 10 and 20 GHz images. 
Finally, the concatenated image of the historical VLA observations (Figure \ref{fig:combinedimage}f) shows a markedly different morphology of M\,31*, but this is likely affected by the poorer resolution and sensitivity in this image. 
We caution that the ``plumes" are only detected at a $\sim$3\,$\sigma$ level. The reality and physical meaning of these substructures will be further discussed in Section \ref{sec:discussion}.

\subsection{Variability}
\begin{figure*}\centering
\epsscale{1}
\plotone{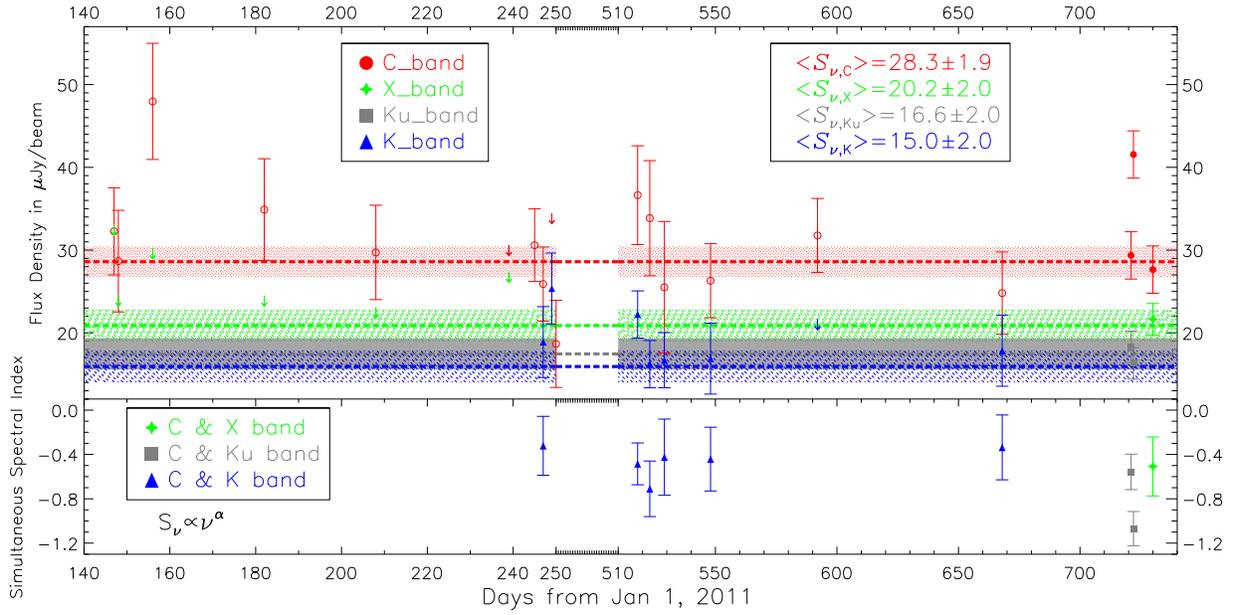}
\caption{{\it Upper panel}: Light curves of M\,31* during the 2011-2012 observing campaign.
Filled circles represent central frequency at 6.0 GHz, while open circles denote 5.0 and 6.6 GHz measurements converted into 6.0 GHz, assuming a spectral index of $-0.45$.
Arrows represent 3-sigma upper limit in the case of non-detections.
The dash lines and corresponding shaded regions show the mean peak flux densities and errors from the concatenated images.
{\it Lower panel}: the instantaneous spectral index derived for the epochs in which M\,31* is detected in dual frequencies.}
\label{fig:curve}
\end{figure*}

We have measured the flux density (or upper limit, in the case of non-detection) of M\,31* in a uniform fashion for both the JVLA and historical VLA images. 
The peak flux density, Gaussian size and integrated flux density are listed for each image (ID) in Table \ref{tab:VLAresults}. 
Figure \ref{fig:curve} presents the light curves of M\,31* derived from the 19 JVLA epochs, with different color-coded symbols representing different bands. 
To facilitate the comparison among the slightly different C-band central frequencies, we have converted the 5.0 or 6.6 GHz flux density into a 6.0 GHz flux density, assuming a spectral index of $\alpha = -0.45$ ($S_\nu \propto \nu^\alpha$; sec Section \ref{subsec:spec}).
From the C-band light curve (red circles), which covers all epochs, we identify two significant variations: the peak flux density reached a maximum of $48.6\pm8.0$ $\mu$Jy/beam on Jun 05, 2011, having increased by $\sim$71\% ($\sim$2\,$\sigma$ significance) in about eight days; 
On December 22, 2012, the peak flux density increased by $\sim$44\% ($\sim$3\,$\sigma$ significance) within just six hours.
Otherwise the C-band light curve appears flat, exhibiting a modest rms fluctuation of $\sim15$\% about its mean ($28.3\pm1.9$ $\mu$Jy/beam).
The light curves at the other bands are much more sparse and show no significant variability, either.
The mean peak flux densities of M\,31* are $20.2\pm2.0$, $16.6\pm2.0$ and $15.0\pm2.0$ $\mu$Jy/beam, derived from the concatenated images at 10, 15 and 20 GHz, respectively.

 More quantitatively, we calculate for each band the variability index \citep{2011ApJ...740...66P}, 
\begin{equation}
VI = \frac{1}{N}\sum_i\left|\frac{S_i-\bar{S}}{\bar{\sigma}}\right|,
\end{equation}
where $S_i$ the flux density at the $i$th epoch, $\bar{S}$ the mean flux density, and $\bar{\sigma}$ the mean error over $N$ epochs. 
A source is considered variable in the case of $VI$ $>$ 1.0, i.e., its flux density varies more than the mean measurement error. 
We find that in the C-band $VI \approx 1.1$, whereas $VI < 1.0$ in the other bands. 
This reinforces our conclusion that during 2011-2012, M\,31* shows marginal variability at 6 GHz, but is consistent with no significant variability at higher frequencies, due partially to the limited coverage.

A comparison with the historical VLA observations allows to probe the long-term variability in M\,31*. 
We measure a mean peak flux density of $60.0\pm10.0$ $\mu$Jy/beam at 5 GHz between 2002-2005, which is consistent with that reported in Garcia et al.~(2010). 
Compared to the mean value of $28.3\pm1.9$ $\mu$Jy/beam at 6 GHz between 2011-2012, this indicates a $\sim$50\%  ($\sim$3\,$\sigma$ significance) decrease over about a decade.
We note that the 5 GHz flux density of our calibrator, 3C48, exhibits no significant variability (to within 0.2\%) in the mean time.
A long-term decrease is also evident at a higher frequency: Crane et al. (1992, 1993) measured 8.4 GHz flux densities of $28 \pm 5$ and $39 \pm 5$ $\mu$Jy, while our observations set a 3\,$\sigma$ upper limit of 22 $\mu$Jy at 8.5 GHz.

\subsection{Spectral Index} \label{subsec:spec}
\begin{figure*}\centering
\epsscale{1}
\includegraphics[width=0.8\textwidth,angle=0]{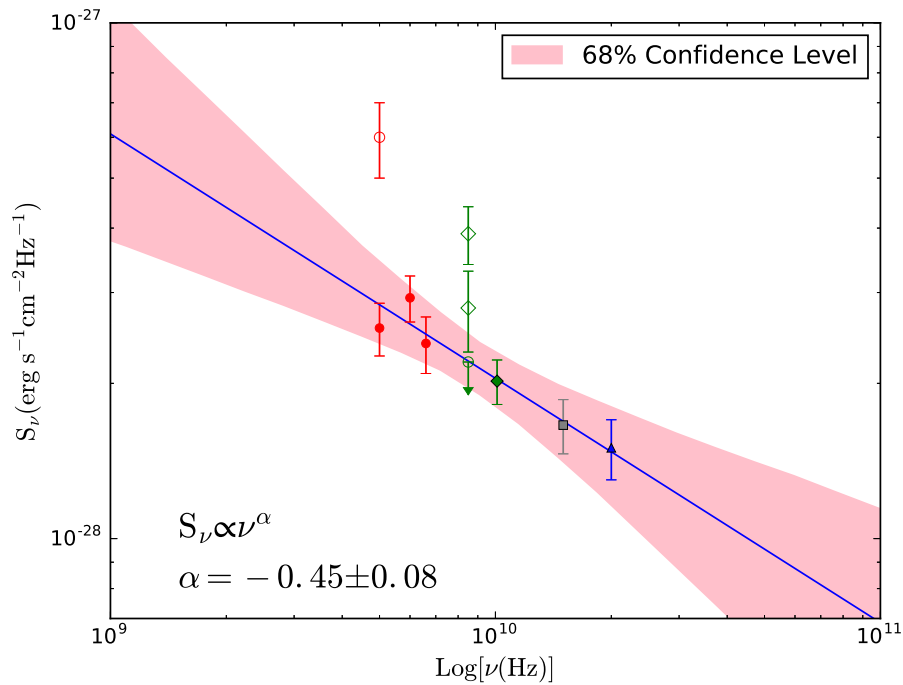}
\vskip3.5cm
\caption{The radio spectral energy distribution of M\,31*. 
The color-coded filled symbols denote the mean peak flux densities, as measured from the concatenated JVLA images (red for C band, green for X band, gray for Ku band and blue for K band). 
The solid line and the shaded region denote the best-fit power-law and its uncertainty, fitted to the filled symbols only.
For comparison, the 3\,$\sigma$ upper limit at 8.5 GHz is marked by an arrow, along with open symbols showing historical VLA measurements at 5 GHz (this work) and 8.5 GHz (Crane et al.~1992, 1993).
}
\label{fig:SED}
\end{figure*}

In 9 of the 17 dual-frequency epochs, M\,31* is detected (i.e., $>$ 3\,$\sigma$) in both frequencies.
We derive a quasi-instananeous spectral index for each of these 9 epochs and find values ranging from $-1.09\pm0.16$ to $-0.31\pm0.28$ (see lower panel of Figure \ref{fig:curve}). 
We also obtain a time-averaged spectral index over the longest possible frequency range (Figure \ref{fig:SED}), based on the peak flux densities at 5, 6.0, 6.6, 10, 15 and 20 GHz from the concatenated images. 
This leads to $\alpha = -0.45\pm0.08$ (if instead the integrated flux densities were adopted, the resultant spectral index is $\alpha = -0.44\pm0.10$), which is consistent with synchrotron radiation. 
The upper limit at 8.5 GHz is compatible with this average spectrum. 
Notably, the average spectrum predicts a flux density of 12.5 $\mu$Jy at 30 GHz and 11.0 $\mu$Jy at 40 GHz, inviting future JVLA observations at these higher frequencies.

\section{Discussion} \label{sec:discussion}
Our JVLA observations reveal unprecedented details about the radio emission from M\,31*. This provides important clues to its otherwise underluminous radiation, and in turn sheds light to the nature of LLAGNs in general, as addressed below.

\subsection{Temporal behavior: LLAGN in context}

The 19 C-band observations taken during 2011-2012 provide insights on the variability in M\,31* arising from a physical scale of $\lesssim10^5 R_{\rm Sch}$. 
In particular, the rms fluctuation is found to be only modest ($\sim$15\%) on weekly to monthly timescales. This is comparable to that in Sgr A* ($\sim17$\% at 5 GHz; Zhao et al.~1989). 
It is often considered that interstellar scintillation contributes substantially to the observed variability in Sgr A* at low frequencies ($\lesssim$ 5 GHz; e.g., \citealp{2006ApJ...641..302M}). Since M\,31* has a much lower line-of-sight column density ($\lesssim 10^{21}{\rm~cm^{-2}}$; Li et al.~2009), the 15\% fluctuation can be considered intrinsic, which might be due to a stochastic accretion rate or plasma instability in the course of jet propagation.   
It remains to be explored whether such fluctuations are typical of LLAGNs.
Significant variations are also caught on short timescales (Section 3.2). In particular, the $\sim$40\% increase in the 6 GHz flux density, which occurred within 6 hrs on December 22, 2012, places a strong constraint on the size of the emitting region, $\lesssim16$ $R_{\rm Sch}$. The amplitude and physical scale of this event are similar to some of the radio flares in Sgr A*. 
We point out, however, the 15 GHz flux density measured at the same epochs showed no significant variability (Figure \ref{fig:curve}).

The comparison between the JVLA and historical VLA measurements further reveals that M\,31* has decreased its mean flux density by about 50\% over about a decade (Section 3.2; Figure 3). 
On such timescales, M\,31* might have experienced a substantial adjustment in its accretion rate, leading to a subsequent change in the jet power. 
Interestingly, the mean X-ray flux of M\,31* had shown an opposite trend in the past decade. 
M\,31* was essentially undetected in any {\it Chandra} observation taken before 2006 (with a mean X-ray luminosity $\lesssim 10^{36}{\rm~erg~s^{-1}}$), but alleviated its X-ray emission by a factor of $\sim$5 on average, apparently after an outburst on January 6, 2006 (Li et al.~2011).
We note that such a trend has not been found in the X-ray emission from Sgr A*, although the two most underluminous LLAGNs both currently exhibit flaring X-ray emission. 
A long-term decrease in the radio flux of Sgr A* has not been reported, either.
Unfortunately, there was no useful VLA observation of M\,31* between March 2005 and April 2011, i.e., the period during which the mean radio flux presumably experienced a substantial (and perhaps abrupt) drop.  
Whether there is a causal relation between the increase in X-rays and decrease in radio remains an intriguing possibility to investigate. We defer such a study, to be assisted with standard ADAF-jet models (Yuan et al. 2009), to a future work.

To our knowledge, long-term radio variability has been reported for few LLAGNs, due partially to the lack of extensive monitoring observations.  
One exception is M\,81*, an archetype LLAGN with a black hole mass of $7\times10^7{\rm~M_{\sun}}$ and a bolometric Eddington ratio of $\sim10^{-5}$ \citep{2003AJ....125.1226D}.
\cite{2011A&A...533A.111M} reported a long-term radio (5.0 and 8.4 GHz) flare in M\,81* from 1997 to 2001, which climbed to a peak more than twice the quiescent level. This suggests that decade-long monitoring observations be crucial to obtaining a clear picture of the variability in LLAGNs.


\subsection{Spectrum and morphology: insights on the putative jet}
We have derived an average spectral index of $\alpha\approx-0.45\pm0.08$ over the available frequency range of 5-20 GHz (Figure \ref{fig:SED}).
While none of our 19 epochs covers more than two frequencies, this average slope is expected to represent the true slope, in view of the very modest variability among these epochs. Indeed, the six epochs with a C/K combination show an instantaneous spectral index ranging between $-0.31\pm0.28$ to $-0.72\pm0.26$.
This relatively steep spectrum is to be contrasted with that of Sgr A*, 
which is observed to be $\alpha \approx 0.2-0.3$ over a similar frequency range \citep{1998ApJ...499..731F}.
In the case of M\,81*, in which substructures are clearly resolved,
\cite{2004ApJ...615..173B} measured~$\alpha \ga0$ near the core and $\alpha\approx-0.7$ in the apparently one-sided jet at $\sim1$ milli-arcsec ($\sim 3\times10^3$ times the Schwarzschild radius of M\,81*) away from the core. 
These are understood in the standard jet models (e.g., \citealp{2000A&A...362..113F, 2010LNP...794..143M}), which predict a negative gradient in the spectral index along the jet path as the result of decreasing synchrotron opacity. 
In this regard, the spectral index of M\,31* ($\sim$-0.45) suggests that the observed radio emission over 5-20 GHz is dominated by the optically-thin part of the jet.
We note that the physical scale corresponding to the synthesized beam of our JVLA images, is $\sim (2-7)\times10^{4}R_{\rm Sch}$. This places an upper limit in the distance of the emitting region from the black hole. The short-timescale variability might take place much closer to the black hole (Section 4.1).

It is natural to look for a trace of the putative jet on larger scales. 
In this regard, the three ``plumes" seen primarily in the concatenated JVLA C-band image (Figure 1a and e) deserve some remarks.
These faint ($\sim$3\,$\sigma$) features are the only hint for an extended component around an otherwise unresolved core.
Their combined morphology, however, is not easily compatible with a canonical bipolar jet/outflow.   
On the other hand, the vicinity of M\,31* on a scale of few arcsecs is known to be devoid of cold gas and young, massive stars (Li et al.~2009). 
Therefore, it is also unlikely that these ``plumes" arise from circumnuclear free-free emission. 
We speculate, on the basis of their combined morphology, the ``plumes" might be tracing the synchrotron radiation from a hot wind launched from the inner part of the accretion flow of M\,31*. Such a wind/outflow is a generic prediction by the theories and numerical simulations of hot accretion flows in LLAGNs (Blandford \& Begelman 1999; Yuan, Bu \& Wu 2012; Narayan et al. 2012; Yuan et al. 2015). 
These theoretical works have gained increasing observational support, from LLAGNs (e.g., Wang et al. 2013; Tombesi et al. 2014) to the hard state of black hole X-ray binaries (Homan et al. 2016).
The predicted winds have a wide opening angle as compared to the highly collimated jet, as they   
originate from the corona region of the hot accretion flow, where frequent magnetic reconnection should occur and can efficiently accelerate electrons (Yuan et al. 2009b).  
Therefore, relativistic electrons should be present in the winds and emit radio emission via synchrotron radiation.
Recently, we detected the radio counterpart of the central SMBH in M\,32 (named M\,32*), a dwarf elliptical companion of M\,31, using deep JVLA C-band observations (Yang et al.~2015). M\,32* has a 6 GHz flux density similar to that of M\,31*, and also shows a pc-scale, extended morphology marginally resolved by the synthesized beam.
This source could be another case of synchrotron radiation from the wind of a LLAGN.
A more meaningful test of our speculation awaits further work.

\section{Summary}
Thanks to the unprecedented sensitivity afforded by our JVLA observations, we have significantly advanced our knowledge about the radio properties of M\,31*, one of the nearest and most underluminous SMBHs. Our main results include:

\begin{enumerate}

\item Detection of M\,31* at 10, 15 and 20 GHz for the first time; 

\item Constraints on the short- and long-term variability in M\,31*. In particular, a curious decrease of the mean flux density, by $\sim$50\%, is found between the JVLA and historical VLA observations separated by nearly a decade, which might be associated with an observed increase in the mean X-ray flux of M\,31*;  

\item Measurement of the spectral index covering 5-20 GHz. The relatively steep spectrum ($\alpha \approx -0.45$) is best interpreted as dominated by the optically-thin part of a putative jet, which is located at no more than a few thousand Schwarzschild radii from the black hole. However, no clear evidence is provided in our images for an extended component tracing this putative jet. 

\end{enumerate}

These findings clearly invite further high-resolution interferometric observations, as well as dedicated numerical models, to deepen our understanding of the formation and evolution of jets and outflows from M\,31*. 

\begin{acknowledgements}
The National Radio Astronomy Observatory is a facility of the National Science Foundation operated under cooperative agreement by Associated Universities, Inc.
This work is supported by the National Natural Science Foundation of China under grants 11473010, 11573051 and 11633006.
We thank Mike Garcia for the effort in preparing Project SD0487.
Z.L. acknowledges support from the Recruitment Program of Global Youth Experts.
F.Y. acknowledges support from the Key Research Program of Frontier Sciences of CAS
(No. QYZDJ-SSW-SYS008) and the Ministry of Science and Technology
of China (No. 2016YFA0400704).
\end{acknowledgements}

\begin{landscape}
\begin{deluxetable}{cccccccccccc}
\tabletypesize{\scriptsize}
\tablecaption{Observation Log and M\,31* Measurements in Individual Images}
\tablewidth{0pt}
\tablehead{
\colhead{}&
\colhead{}&
\colhead{}&
\colhead{}&
\colhead{}&
\colhead{}&
\colhead{}&
\colhead{}&
\colhead{}&
 \multicolumn{3}{c}{IMFIT Gaussian}\\ 
\cline{10-12}
\colhead{ID}&
\colhead{Project}&
\colhead{Date}&
\colhead{$\nu$}&
\colhead{Time}&
\colhead{Config.}&
\colhead{Bandwidth}&
\colhead{Beam size}&
\colhead{RMS}&
\colhead{$I_{{\rm peak}}$}&
\colhead{Image component size}&
\colhead{$S_{{\rm int}}$}\\
\colhead{}&
\colhead{}&
\colhead{}&
\colhead{(GHz)}&
\colhead{(hr)}&
\colhead{}&
\colhead{(MHz)}&
\colhead{($\arcsec\times\arcsec, \degr$)}&
\colhead{($\mu$Jy/beam)}&
\colhead{($\mu$Jy/beam)}&
\colhead{($\arcsec\times\arcsec, \degr$)}&
\colhead{($\mu$Jy)}\\
\colhead{(1)}&
\colhead{(2)}&
\colhead{(3)}&
\colhead{(4)}&
\colhead{(5)}&
\colhead{(6)}&
\colhead{(7)}&
\colhead{(8)}&
\colhead{(9)}&
\colhead{(10)}&
\colhead{(11)}&
\colhead{(12)}\\
}
\startdata
1&11A-178& 2011 May 27& 5.0& 2.5& BnA& 256& $1.52\times0.67$, -68&6.0&$34.9$&$1.58\times0.78$, 113&$42.1\pm5.5$\\
2&11A-178& 2011 May 27& 8.5&  2.5& BnA& 256& $0.91\times0.42$, -66& 11.0&$< 33.0$ & - & -\\
3&11A-178& 2011 May 28& 5.0&  2.5& BnA& 256& $1.52\times0.62$, -68&7.0 &$30.8$&$1.66\times0.91$, 129&$49.0\pm9.6$\\
4&11A-178& 2011 May 28& 8.5&2.5  & BnA& 256& $0.95\times0.38$, -65&8.0 &$< 24.0$&-&-\\
5&11A-178& 2011 Jun 05&  5.0& 2.5 & BnA& 256& $0.71\times0.50$, -87&8.0 &$52.8$&$0.79\times0.53$, 96&$62.2\pm7.9$\\
6&11A-178& 2011 Jun 05&  8.5& 2.5 & BnA& 256& $0.45\times0.30$, 89&10.0 &$< 30.0$&-&-\\
7&11A-178& 2011 Jul 01&   5.0&2.5  & A& 256& $0.50\times0.45$, -74&7.0 &$37.9$&$0.50\times0.45$, -74&$37.9\pm7.0$\\
8&11A-178& 2011 Jul 01&   8.5& 2.5 & A& 256& $0.30\times0.27$, -73&8.0 &$< 24.0$&-&-\\
9&11A-178& 2011 Jul 27&   5.0& 2.5 & A& 256& $0.59\times0.44$, -78& 6.5&$32.0$&$0.59\times0.44$, -78&$32.0\pm6.5$\\
10&11A-178& 2011 Jul 27&   8.5& 2.5 & A& 256& $0.38\times0.27$, -71& 7.5&$< 23.0$&-&-\\
11&11A-178& 2011 Aug 27& 5.0& 2.5 & A& 256& $0.46\times0.41$, -6& 11.0&$< 33.0$&-&-\\
12&11A-178& 2011 Aug 27& 8.5& 2.5 & A& 256& $0.28\times0.24$, 12& 9.0&$< 27.0$&-&-\\
13&SD0487&  2011 Sep 02&   5.0& 6 & A& 256& $0.45\times0.43$, -4&5.0 &$33.0$&$0.62\times0.46$, 113&$48.6\pm5.7$\\
14&11A-137& 2011 Sep 04&   6.6& 1 & A& 2048& $0.36\times0.33$, -73&4.5 &$24.4$&$0.36\times0.33$, -73&$24.4\pm4.5$\\
15&11A-137& 2011 Sep 04&   20& 4.5 & A& 2048& $0.12\times0.11$, -5&4.5&$18.1$&$0.12\times0.11$, -5&$18.1\pm4.5$\\
16&11A-137& 2011 Sep 06&   6.6& 1 & A& 2048& $0.37\times0.33$, -64&11.0 &$< 33.0$&-&-\\
17&11A-137& 2011 Sep 06&   20& 4.5 & A& 2048& $0.11\times0,10$, -10&4.5 &$24.9$&$0.11\times0.10$, -10&$24.9\pm4.5$\\
18&SD0487&  2011 Sep 07&   5.0& 6 & A& 256& $0.57\times0.44$, -40&6.0 &$19.4$&$0.75\times0.66$, 159&$38.2\pm6.6$\\
19&SD0487&  2012 Jun 01&   6.6& 1.5 & B& 2048& $1.20\times1.08$, -86&6.0 &$35.2$&$1.20\times1.08$, -86&$35.2\pm6.0$\\
20&SD0487&  2012 Jun 01&   20& 4.5 &  B& 2048& $0.35\times0.35$, 12&3.0 &$21.6$&$0.42\times0.33$, 94&$24.2\pm1.2$\\
21&SD0487&  2012 Jun 06&   6.6& 1.5 & B& 2048& 1.35$\times$1.11, 53& 7.0&$32.4$&$1.59\times1.44$, 97&$49.0\pm11.0$\\
22&SD0487&  2012 Jun 06&   20& 4.5 &  B& 2048& 0.40$\times$0.35, 38& 3.0&$15.3$&$0.40\times0.35$, 38&$15.3\pm3.0$\\
23&SD0487&  2012 Jun 12&   6.6& 1.5 & B& 2048& 1.22$\times$1.18, 80& 8.0&$24.0$&$2.42\times1.46$, 168&$58.7\pm5.2$\\
24&SD0487&  2012 Jun 12&   20& 4.5 &  B& 2048& 0.39$\times$0.35, -26&3.5 &$15.8$&$0.39\times0.35$, -26&$15.8\pm7.0$\\
25&SD0487&  2012 Jul 01&   6.6& 1.5 & B& 2048& 1.23$\times$1.05, -89&4.5 &$24.8$&$1.50\times0.98$, 55&$28.3\pm4.7$\\
26&SD0487&  2012 Jul 01&   20& 4.5 &  B& 2048& 0.35$\times$0.34, -54& 4.5&$16.0$&$0.50\times0.31$, 24&$20.4\pm4.5$\\
27&SD0487&  2012 Aug 14&   6.6&1.5 & B& 2048& 1.24$\times$1.10, 76& 4.5&$30.3$&$1.24\times1.10$, 176&$30.3\pm4.5$\\
28&SD0487&  2012 Aug 14&   20& 4.5 &  B& 2048& 0.37$\times$0.35, -76&7.0 &$< 21.0$&-&-\\
29&SD0487&  2012 Oct 29&   6.6& 1.5 & A& 2048& 0.39$\times$0.33, -80& 5.0&$23.3$&$0.60\times0.48$, 29&$52.6\pm5.7$\\
30&SD0487&  2012 Oct 29&   20& 4.5 &  A& 2048& 0.16$\times$0.13, -22& 4.5&$17.0$&$0.19\times0.15$, 158&$22.4\pm4.6$\\
31&12B-002&  2012 Dec 22&   6.0& 0.9 &  A& 4000& 0.36$\times$0.36, 31&3.0 &$29.1$&$0.45\times0.32$, 100&$31.1\pm3.4$\\
32&12B-002&  2012 Dec 22&   15.0& 2.7 &  A& 6000& 0.16$\times$0.14, 35& 2.0&$17.5$&$0.19\times0.16$, 26&$23.2\pm2.8$\\
33&12B-002&  2012 Dec 22&   6.0& 0.9 & A& 4000& 0.45$\times$0.36, 80&3.0 &$41.9$&$0.53\times0.48$, 102&$65.9\pm5.1$\\
34&12B-002&  2012 Dec 22&   15.0& 2.7 &  A& 6000& 0.18$\times$0.17, 73&2.0 &$15.4$&$0.18\times0.18$, 96&$16.9\pm1.7$\\
35&12B-002&  2012 Dec 30&   6.0& 0.9 &  A& 4000& 0.39$\times$0.36, -74&3.0 &$27.3$&$0.39\times0.36$, -74&$27.3\pm3.7$\\
36&12B-002&  2012 Dec 30&   10.0& 2.7 &  A&4000 & 0.25$\times$0.23, -20&2.0 &$21.0$&$0.25\times0.25$, 11&$22.7\pm2.2$\\
\hline
\hline
37&AJ0289&2002 Jul 05&5&8&B&100&$1.52\times1.46$, 83&7.0&$25.8$&$1.52\times1.46$, 83&$25.8\pm7.0$\\
38&AJ0289&2002 Jul 06&5&8&B&100&$1.52\times1.46$, 85&7.0&$32.8$&$1.52\times1.46$, 85&$32.8\pm7.0$\\
39&AJ0289&2002 Jul 29&5&8&B&100&$1.53\times1.48$, -88&7.0&$48.7$&$1.53\times1.48$, -88&$48.7\pm7.0$\\
40&AJ0289&2002 Aug 12&5&2&B&100&$1.69\times1.43$, -80&12.0&$37.8$&$1.69\times1.43$, -80&$37.8\pm12.0$\\
41&TLS17&2002 Aug 14&5&1&B&100&$1.53\times1.40$, -53&15.0&$66.7$&$1.53\times1.40$, -53&$66.7\pm15.0$\\
42&AS0819&2004 Oct 27/28&	5&6.3&A&		100&$0.66\times0.44$, -71&18.0&$65.0$&$0.72\times0.31$, 51&$65.0\pm18.0$\\
43&S6314& 2004 Dec 06/07&   5& 12  &A&     100&   $0.64\times0.51$, 68  &8.0&$54.4$&$0.64\times0.51$, 68& $54.4\pm8.0$\\
44&S6314& 2004 Dec 27/28&   5& 12 &A&     100&   $0.56\times0.44$, -88  &9.0&$54.0$&$0.56\times0.44$, -88&$54.0\pm9.0$\\
45&S6314& 2005 Jan 28/29&   5& 12  &AB&   100&   $1.27\times0.67$, 80&9.0&$65.0$&$1.27\times0.67$, 80&$65.0\pm9.0$\\
46&S6314& 2005 Feb 21/22&   5& 12  &B&     100&   $1.83\times1.41$, -79 &8.0&$67.9$&$1.83\times1.41$, -79&$67.9\pm8.0$\\
\enddata
\vspace{-0.6cm}
\tablecomments{(1) Image ID; (2) Project ID; (3) Date of observation; (4) The central frequency; (5) Total integration time; (6) Array and configuration; (7) Total bandwidth; (8) Synthesized beam FWHM of major and minor axes, and position angle of the major axis; (9) Image RMS level; (10) Peak flux density of M\,31*; (11) major axis, minor axis and position angle of the fitted elliptical Gaussian; (12) integrated flux density.}
\label{tab:VLAresults}
\end{deluxetable}
\end{landscape}

\clearpage

\begin{landscape}
\begin{deluxetable}{cccccccccc}
\tabletypesize{\scriptsize}
\tablecaption{Mean Flux Densities of M\,31*}
\tablewidth{0pt}
\tablehead{
\colhead{}&
\colhead{}&
\colhead{}&
\colhead{}&
\colhead{}&
\colhead{}&
 \multicolumn{3}{c}{IMFIT Gaussian}\\ 
 \cline{8-10}
 Reference image &ID&Config.&$\nu$&Time&Beam size&$RMS$&$I_{\rm peak}$&Image component size&$S_{\rm int}$\\
&&&(GHz)&(hr)&($\arcsec\times\arcsec$, $\degr$)&$(\mu$Jy/beam)&$(\mu$Jy/beam)&($\arcsec\times\arcsec$, $\degr$)&($\mu$Jy)\\
(1)&(2)&(3)&(4)&(5)&(6)&(7)&(8)&(9)&(10)\\
}
\startdata
\multirow{4}{*}{Fig. 1(a)}&1, 3, 5, 7, 9, 11, 13, &\multirow{4}{*}{A, B, BnA}&\multirow{4}{*}{6.0}&\multirow{4}{*}{40.7}&\multirow{4}{*}{$0.58\times0.50$, -78}&\multirow{4}{*}{1.9}&\multirow{4}{*}{$28.3$}&\multirow{4}{*}{$0.64\times0.57$, 125}&\multirow{4}{*}{$35.5\pm4.4$}\\
&14, 16, 18, 19, 21, \\
&23, 25, 27, 29, 31,\\
 &33, 35&&&&&\\

Fig. 1(b)&36								&A&10&2.7&$0.25\times0.23$, -14&2.0&$20.2$&$0.26\times0.26$, 39&$23.9\pm3.5$\\
Fig. 1(c)&32, 34							&A&15&5.4&$0.17\times0.16$, 46&2.0&$16.6$&$0.18\times0.17$, 19&$20.7\pm2.1$\\
\multirow{2}{*}{Fig. 1(d)}&15, 17, 20, 22,&\multirow{2}{*}{A, B}&\multirow{2}{*}{20}&\multirow{2}{*}{27}&\multirow{2}{*}{$0.25\times0.24$, -3}&\multirow{2}{*}{2.0}&\multirow{2}{*}{$15.0$}&\multirow{2}{*}{$0.27\times0.22$, 9}&\multirow{2}{*}{$17.0\pm3.4$}\\
&24, 26, 28, 30&\\
\multirow{2}{*}{Fig. 1(e)}&7, 9, 11, 13, 14, 16,&\multirow{2}{*}{A}&\multirow{2}{*}{6.0}&\multirow{2}{*}{22.2}&\multirow{2}{*}{$0.41\times0.38$, -81}&\multirow{2}{*}{1.2}&\multirow{2}{*}{$28.7$}&\multirow{2}{*}{$0.51\times0.42$, 114}&\multirow{2}{*}{$39.7\pm3.2$}\\
&18, 29, 31, 33, 35&&&&\\
Fig. 1(f)&37-46							&A, B&5&81.3&$0.88\times0.66$, 80&10.0&$60.0$&$0.79\times0.65$, 87&$62.0\pm7.6$\\
\enddata
\vspace{-0.6cm}
\tablecomments{(1) Concatenated image in Figure \ref{fig:combinedimage}; (2) ID as in Table 1; (3) Configuration; (4) Central frequency; (5) Total integration time; (6) Synthesized beam size, including major and minor axis, and position angle of the major axis; (7) Image RMS noise; (8) Peak flux densities of M\,31*; (9) Major and minor axes, and position angle of the fitted elliptical Gaussian; (10) Integrated flux density.}
\label{tab:concat}
\end{deluxetable}
\end{landscape}

\clearpage

\acknowledgments

\end{document}